\def\ppbar{$p\overline{p} $}            
\def\D0{D\O}                            
\def\ipb{pb$^{-1}$}                     
\def\pt{$p_T$}                          
\def\et{$E_T$}                          
\def\gevcc{GeV/c$^2$}                   
\def\gevc{GeV/c}                        
\def\gev{GeV}                           
\def\iso{$\cal{I}$}                     
\def\wino{$\widetilde W$}               
\def\zino{$\widetilde Z$}               
\def\met{\mbox{${\hbox{$E$\kern-0.6em\lower-.1ex\hbox{/}}}_T$ }} 
\def\d0draft{}

\def\err#1#2#3 {{\it Erratum} {\bf#1},{\ #2} (19#3)}
\def\ib#1#2#3 {{\it ibid.} {\bf#1},{\ #2} (19#3)}
\def\nc#1#2#3 {Nuovo Cim. {\bf#1} ,#2(19#3)}
\def\nim#1#2#3 {Nucl. Instr. Meth. {\bf#1},{\ #2} (19#3)}
\def\np#1#2#3 {Nucl. Phys. {\bf#1},{\ #2} (19#3)}
\def\pl#1#2#3 {Phys. Lett. {\bf#1},{\ #2} (19#3)}
\def\prev#1#2#3 {Phys. Rev. {\bf#1},{\ #2} (19#3)}
\def\prl#1#2#3 {Phys. Rev. Lett. {\bf#1},{\ #2} (19#3)}
\def\rmp#1#2#3 {Rev. Mod. Phys. {\bf#1},{\ #2} (19#3)}
\def\zp#1#2#3 {Zeit. Phys. {\bf#1},{\ #2} (19#3)}

\documentstyle[preprint,epsf,eqsecnum,aps,floats]{revtex}  
\begin{document}

\def\theequation{\arabic{equation}}
\def\metns{${\hbox{$E$\kern-0.6em\lower-.1ex\hbox{/}}}_T$}


\preprint{FERMILAB Pub-95/385-E}
\title{Search for \wino$_1$\zino$_2$\ Production via Trilepton Final \\
States in \ppbar\ Collisions at $\sqrt{s} = 1.8$ TeV }

%
\author{
S.~Abachi,$^{14}$
B.~Abbott,$^{28}$
M.~Abolins,$^{25}$
B.S.~Acharya,$^{44}$
I.~Adam,$^{12}$
D.L.~Adams,$^{37}$
M.~Adams,$^{17}$
S.~Ahn,$^{14}$
H.~Aihara,$^{22}$
J.~Alitti,$^{40}$
% ...
\\
\vskip 0.50cm
\centerline{(D\O\ Collaboration)}
\vskip 0.50cm
}
\address{
\centerline{$^{1}$Universidad de los Andes, Bogot\'{a}, Colombia}
\centerline{$^{2}$University of Arizona, Tucson, Arizona 85721}
\centerline{$^{3}$Boston University, Boston, Massachusetts 02215}
\centerline{$^{4}$Brookhaven National Laboratory, Upton, New York 11973}
\centerline{$^{5}$Brown University, Providence, Rhode Island 02912}
\centerline{$^{6}$Universidad de Buenos Aires, Buenos Aires, Argentina}
\centerline{$^{7}$University of California, Davis, California 95616}
\centerline{$^{8}$University of California, Irvine, California 92717}
\centerline{$^{9}$University of California, Riverside, California 92521}
\centerline{$^{10}$LAFEX, Centro Brasileiro de Pesquisas F{\'\i}sicas,
                  Rio de Janeiro, Brazil}
\centerline{$^{11}$CINVESTAV, Mexico City, Mexico}
\centerline{$^{12}$Columbia University, New York, New York 10027}
\centerline{$^{13}$Delhi University, Delhi, India 110007}
\centerline{$^{14}$Fermi National Accelerator Laboratory, Batavia,
                   Illinois 60510}
\centerline{$^{15}$Florida State University, Tallahassee, Florida 32306}
\centerline{$^{16}$University of Hawaii, Honolulu, Hawaii 96822}
\centerline{$^{17}$University of Illinois at Chicago, Chicago, Illinois 60607}
\centerline{$^{18}$Indiana University, Bloomington, Indiana 47405}
\centerline{$^{19}$Iowa State University, Ames, Iowa 50011}
\centerline{$^{20}$Korea University, Seoul, Korea}
\centerline{$^{21}$Kyungsung University, Pusan, Korea}
\centerline{$^{22}$Lawrence Berkeley National Laboratory and University of
                   California, Berkeley, California 94720}
\centerline{$^{23}$University of Maryland, College Park, Maryland 20742}
\centerline{$^{24}$University of Michigan, Ann Arbor, Michigan 48109}
\centerline{$^{25}$Michigan State University, East Lansing, Michigan 48824}
\centerline{$^{26}$Moscow State University, Moscow, Russia}
\centerline{$^{27}$University of Nebraska, Lincoln, Nebraska 68588}
\centerline{$^{28}$New York University, New York, New York 10003}
\centerline{$^{29}$Northeastern University, Boston, Massachusetts 02115}
\centerline{$^{30}$Northern Illinois University, DeKalb, Illinois 60115}
\centerline{$^{31}$Northwestern University, Evanston, Illinois 60208}
\centerline{$^{32}$University of Notre Dame, Notre Dame, Indiana 46556}
\centerline{$^{33}$University of Oklahoma, Norman, Oklahoma 73019}
\centerline{$^{34}$University of Panjab, Chandigarh 16-00-14, India}
\centerline{$^{35}$Institute for High Energy Physics, 142-284 Protvino, Russia}
\centerline{$^{36}$Purdue University, West Lafayette, Indiana 47907}
\centerline{$^{37}$Rice University, Houston, Texas 77251}
\centerline{$^{38}$Universidade Estadual do Rio de Janeiro, Brazil}
\centerline{$^{39}$University of Rochester, Rochester, New York 14627}
\centerline{$^{40}$CEA, DAPNIA/Service de Physique des Particules, CE-SACLAY,
                   France}
\centerline{$^{41}$Seoul National University, Seoul, Korea}
\centerline{$^{42}$State University of New York, Stony Brook, New York 11794}
\centerline{$^{43}$SSC Laboratory, Dallas, Texas 75237}
\centerline{$^{44}$Tata Institute of Fundamental Research,
                   Colaba, Bombay 400005, India}
\centerline{$^{45}$University of Texas, Arlington, Texas 76019}
\centerline{$^{46}$Texas A\&M University, College Station, Texas 77843}
}

\date{\today}

\maketitle

\begin{abstract}
We have searched for associated production of the lightest
chargino, $\widetilde{W}_1$, and
next-to-lightest neutralino, $\widetilde{Z}_2$, of the Minimal Supersymmetric
Standard
Model in $p\bar{p}$\ collisions at \mbox{$\sqrt{s}$ =\ 1.8\ TeV} using
the \D0 detector at the Fermilab Tevatron collider. Data corresponding to
an integrated luminosity of 12.5$\pm 0.7$\ \ipb\ were examined for events
containing
three isolated leptons. No evidence for $\widetilde{W}_1\widetilde{Z}_2$\ pair
production was found. Limits on
$\sigma(\widetilde{W}_1\widetilde{Z}_2)$Br$(\widetilde{W}_1\rightarrow l\nu
\widetilde{Z}_1)$Br$(\widetilde{Z}_2\rightarrow l\bar{l}\widetilde{Z}_1)$\
are presented.
\end{abstract}
\vspace{0.25in}
\centerline{\it Submitted to the Physical Review Letters}
\vspace{0.25in}


Supersymmetry (SUSY) is a symmetry which relates
bosons and fermions~\cite{SUSY}. Supersymmetric extensions of the
Standard Model (SM) are attractive because they remove
the ``fine tuning" problem
associated with loop corrections to the mass of the Higgs boson and provide
a basis for gauge coupling unification at a high mass scale.
One consequence of these models is the introduction of a SUSY partner
(sparticle) for each SM state.
Every sparticle and SM particle is assigned an internal quantum number called
R-parity.
If R-parity is conserved (as assumed in this analysis), then sparticle states
are produced in pairs and there must be one sparticle which does not decay.
This sparticle is referred to as the Lightest Supersymmetric Particle (LSP).
The SUSY framework which introduces the fewest
additional particles is known as the Minimal Supersymmetric Standard
Model (MSSM)~\cite{MSSM}.  If the requirement is made that SUSY
be a locally invariant gauge symmetry,
the result is a theoretical framework known as supergravity
(SUGRA)~\cite{ARN_NATH}.

In the MSSM and minimal SUGRA there are
two chargino states
(\mbox{$\widetilde{W}_{i,i=1,2}$})
and four neutralino
states (\mbox{$\widetilde{Z}_{i,i=1,4}$}), corresponding to
mixtures of the SUSY partners of the Higgs bosons, $W$\ and $Z$\ bosons, and
the photon.
(In some of the literature an alternate notation
is used: \mbox{$\widetilde{\chi}^{\pm}_{i,i=1,2}$} for charginos and
\mbox{$\widetilde{\chi}^{\circ}_{i,i=1,4}$} for neutralinos.)
In most regions of the SUGRA parameter space not excluded by previous
experiments,
the LSP is the lightest neutralino~\cite{HOWIE_LEP} and thus escapes detection.
The best limits to date on the masses of the
\wino$_1$\ and \zino$_2$\
states come from the LEP experiments~\cite{LEPLIMS}; the current limits are
\mbox{$M_{\widetilde{W}_1} > 45$} \gevcc\ and
\mbox{$M_{\widetilde{Z}_2} > 40-45$} \gevcc.

At \ppbar\ colliders charginos and neutralinos can be produced in pairs, with
\wino$_1$\zino$_2$
pairs having the largest cross section over much of the parameter
space~\cite{HOWIE_TeV}.
Production cross sections
$\cal{O}$(100--10) pb are possible at the Tevatron for \wino$_1$ masses between
45 and
100 \gevcc~\cite{ARNW,LOPEZ}.
The \wino$_1$ can decay into $q\bar{q}'$\ or $l\bar{\nu}$\ plus an LSP, while
the \zino$_2$ can
decay into $q\bar{q}$\ or $l\bar{l}$\ plus an LSP.
The presence of neutrinos or LSP's among the decay products will generally lead
to missing transverse energy (\metns).
The final state consisting of three leptons and \met (and little hadronic
activity)
has few SM backgrounds and is the subject of the present analysis.

\begin{figure}[htbp]
\centerline{\epsfxsize=6in \epsffile{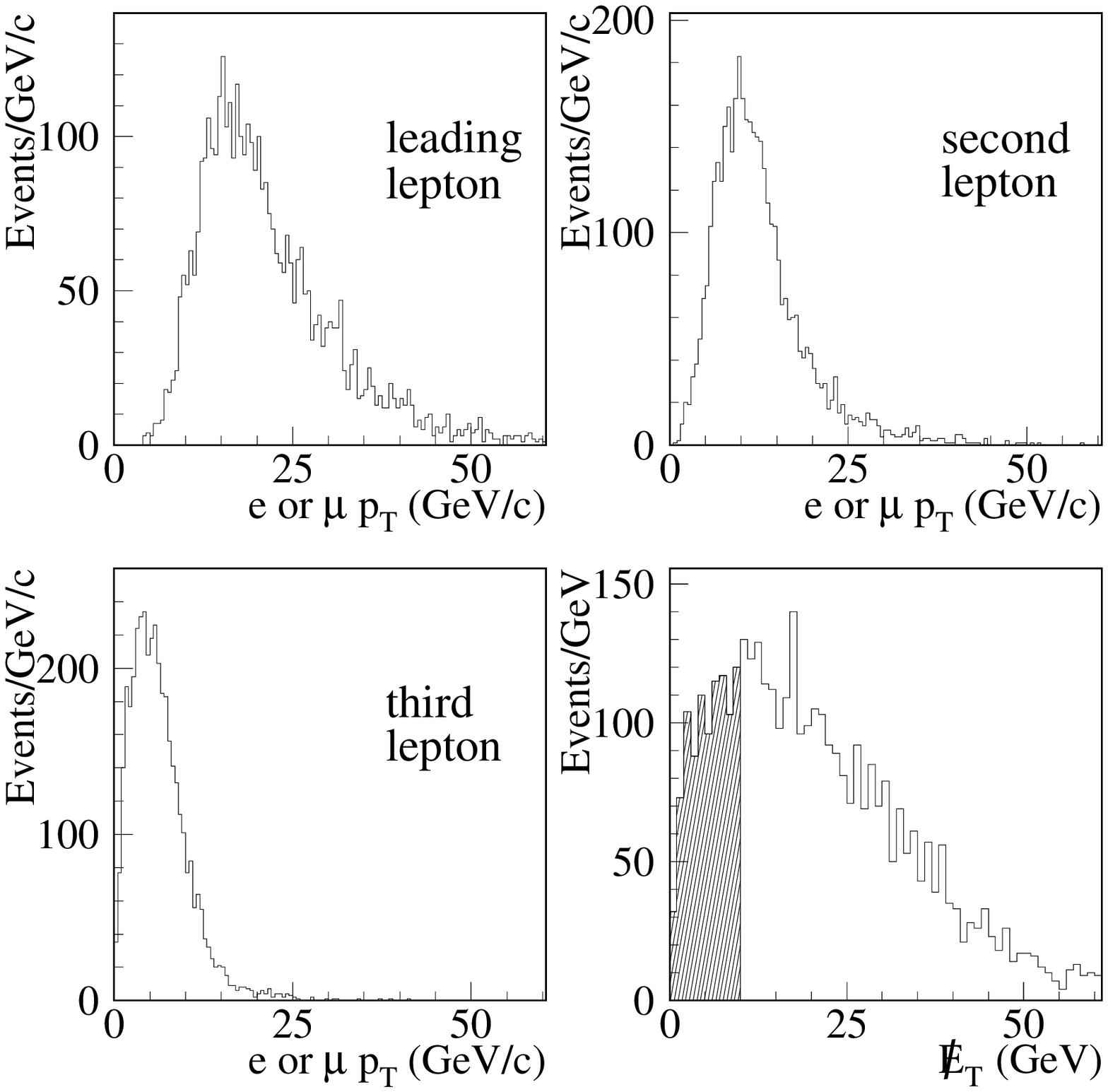}}
\vspace{6in}
\caption{The \pt\ distributions of final state leptons and the \met
distribution
in $\widetilde{W}_1\widetilde{Z}_2 \to 3l$\ events. Events were generated with
$M_{\widetilde{W}_1}
\approx M_{\widetilde{Z}_2} = 56$\ \gevcc.
The shaded area shows the region excluded by the 10 GeV \met\ cut in
the $eee$\ channel.}
\label{fig:ptdistr}
\end{figure}

The spectra of the transverse momenta (\pt) of the final state leptons
can be relatively soft due to the three-body decays of the
\wino$_1$ and \zino$_2$
involving massive non-interacting particles.  Figure~\ref{fig:ptdistr} shows
the expected
\pt\ spectra of the final state leptons as well as the \met distribution
at the physics generator level
for simulated $\widetilde{W}_1\widetilde{Z}_2\to 3l$\ events,
with $M_{\widetilde{W}_1} = 56$\ \gevcc.  These Monte Carlo events follow the
mass relation common to many SUSY models: $M_{\widetilde{W}_1}
\approx M_{\widetilde{Z}_2}
\approx 2 M_{\widetilde{Z}_1}$.

The data used in this analysis were obtained using the \D0 detector at the
Fermilab Tevatron \ppbar\ collider operating at a center of mass
energy of 1.8 TeV. The total integrated luminosity used in this analysis from
the 1992--1993 Tevatron run was $12.5\pm 0.7$ \ipb.

The \D0 detector has three major subsystems: central tracking detectors (with
no central magnetic field), uranium--liquid argon electromagnetic and hadronic
calorimeters, and a muon spectrometer. The detector and data acquisition system
are described in
detail elsewhere\cite{D0NIM}. The central tracking system is used to
identify charged tracks in the pseudorapidity range $|\eta|\leq 3.5$. The
calorimeters provide full angular coverage for  $|\eta|\leq 4.0$,
with transverse segmentation $\Delta\eta\times\Delta\phi = 0.1\times 0.1$,
where $\phi$\ is the azimuthal angle.
The muon system consists of proportional drift chambers and magnetized iron
toroids with coverage extending to $|\eta|\leq 3.3$.

Electrons were identified as calorimeter clusters having at least 90\% of their
energy deposition in the electromagnetic calorimeter, with one or more
tracks pointing to the cluster. Jets were reconstructed from
energy deposition in the calorimeters using a cone algorithm with cone size
${\cal{R}}=\sqrt{\Delta\eta^2+\Delta\phi^2}= 0.5$.
Muon tracks were reconstructed using hits in the muon drift chambers; their
momenta
were calculated from the bend of the tracks in the toroids.

Combinations of single lepton and dilepton triggers were used for the four
final states ($eee$, $ee\mu$, $e\mu\mu$, and $\mu\mu\mu$).
These triggers included:
a single muon with $p_T^{\mu} > 15$\ \gevc;
two muons with $p_T^{\mu} > 3$\ \gevc;
one muon with $p_T^{\mu} > 5$\ \gevc\
plus one electromagnetic cluster with $E_T^{e} > 7$\ \gev;
one electromagnetic cluster with $E_T^{e} > 20$\ \gev;
and two electromagnetic clusters with $E_T^{e} > 10$\ \gev.
The integrated luminosity per channel is given in Table~\ref{tab:AnalCuts}.

Events passing the trigger requirements were selected offline by requiring
three or more reconstructed leptons (electrons or muons) having $E_T^{e} >$\
5 \gev\ or $p_T^{\mu} >$ 5 \gevc, with $|\eta_{e}|<2.5$
or $|\eta_{\mu}| < 1.7$.
There were 2827 events in this initial data sample.
Electrons and muons in these events were then required to pass
the quality cuts described below.

Electrons were required to have transverse and longitudinal shower profiles
consistent with
expectations based on detailed Monte Carlo studies~\cite{topPRD},
to have no more than two tracks pointing to
the calorimeter cluster, and to have an electromagnetic
isolation ${\cal{I}} < 0.15$, where
$ {\cal{I}} = [E_{\text{tot}} - E_{\text{EM}}]/E_{\text{EM}}$,
$E_{\text{tot}}$\ is the total cluster energy inside a cone of
radius ${\cal{R}}=0.4$, and $E_{\text{EM}}$\ is the
electromagnetic energy inside a cone of ${\cal{R}}=0.2$.
For electrons with \et\ between 5 and 10 \gev, the isolation cut was
relaxed to \iso$ < 0.2$\ to increase efficiency.

Muons were required to have a separation from any jet of at least
${\cal{R}}=0.5$,
to be aligned with minimum ionizing energy deposition in at least 50\% of all
calorimeter layers
and in at least 60\% of the hadronic calorimeter layers,
and to have either a matching track in the central detectors or impact
parameters in the
$rz$ (bend) and $xy$ (non-bend) views consistent with the muon having been
produced at the
primary event vertex~\cite{topPRD}.
To reduce cosmic ray background, muons were required to be in time
with the beam crossing and any muon pair having both polar and azimuthal
opening angles
greater than $165^\circ$ was rejected.

There were 19 events 
after these quality cuts.
The following
topological cuts were applied to these events. For the $eee$\ channel, events
were required to have \met$ > 10$\ \gev, with the \met\ reconstructed using
only energy deposited in the calorimeters.
This cut reduced background from $Z/\gamma \rightarrow e^+e^-$\ events with a
third electron from
either a photon conversion (including $\pi^{0} \rightarrow\gamma\gamma$) into
an unresolved
$e^{+}e^{-}$\ pair or a jet which
was reconstructed as an electron.  Since extra material in the forward region
enhances
the photon conversion probability, the data exhibit an excess of electrons
in the forward region while the signal distributions peak in the central
region.
Therefore, a cut was applied in the
$eee$\ and $ee\mu$\ channels to exclude events with more than one electron in
the region $|\eta| > 1.7$.
For the $e\mu\mu$\ and $\mu\mu\mu$\ channels, muon pairs were required to have
an invariant mass greater than 5 \gevcc, which reduced background from
$J/\psi$\ events and the combinatoric background in the reconstruction
of muons.
Table~\ref{tab:AnalCuts} summarizes the effect of the cuts on each of
the channels. We see no candidate events consistent with \wino$_1$\zino$_2$
pair production and subsequent decay into trilepton final states.

Detection efficiencies were determined using a combination of data and Monte
Carlo simulations.  Monte Carlo signal events were
generated using {\small ISAJET}~\cite{ISAJET} and processed with a full
simulation of the \D0 detector
based on the {\small GEANT}~\cite{GEANT} program.
Seven sets of events were generated, with the mass of the \wino$_1$\ varying
from 45 to 100 \gevcc.
Because of the correlation between the masses of the \wino$_1$, \zino$_2$, and
\zino$_1$, efficiencies
can be parametrized as a function of $M_{\widetilde{W}_1}\!$.
These Monte Carlo events were used to determine kinematic and geometric
acceptances only.

Electron identification efficiencies were determined from a set of simulated
single electron
events generated in six \et\ bins between 5 and 25 GeV.  These were overlaid
with minimum bias events
from collider data in order to include the
effects of the underlying event and any noise in the calorimeter on electron
isolation and
shower profile. The results of these studies for high \et\ electrons were
verified
by analyzing a sample of $Z\to ee$\ events~\cite{SOSEBEE} in which one electron
was required to pass all cuts and
the second electron was then used as an unbiased estimator for each cut.

Similarly, muon identification efficiencies were based on $Z \to \mu\mu$\
and $J/\psi \to \mu\mu $\ event samples.  These two sets provided independent
estimates of efficiencies for both high and low \pt\ muons.

\begin{figure}
\centerline{\epsfxsize=6in \epsffile{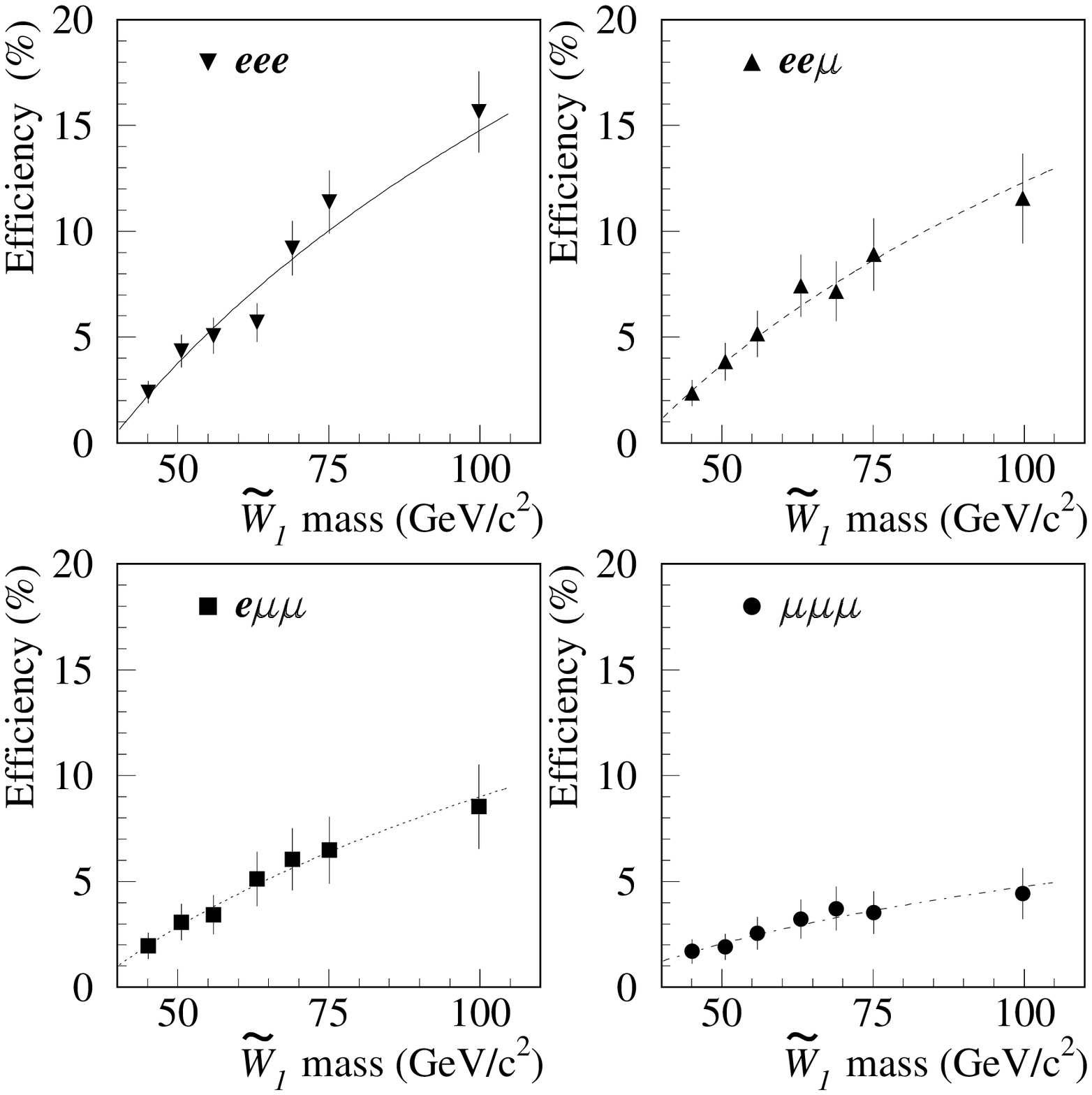}}
\vspace{6in}
\caption{Overall analysis efficiency for each final state as a function of the
mass of the \wino$_1$.}
\label{fig:effs}
\end{figure}

Electron and muon identification efficiencies were parametrized as a function
of the
electron \et\ or muon \pt\ and incorporated with the topological cuts described
above
to determine the overall analysis efficiency for each set of Monte Carlo
signal events. These efficiencies are shown in Fig.~\ref{fig:effs} for each
final state, along with a parametrized fit~\cite{SOSEBEE}, as a function of the
\wino$_1$~mass.

Backgrounds were estimated from data whenever possible, supplemented with Monte
Carlo simulations.
Standard Model processes which produce three or more isolated leptons, such
as vector boson pair production and semileptonic decays in heavy flavor
production,
are expected to yield less than 0.1 event in any channel.
Thus the primary sources of background are single lepton and dilepton events
with one
or more spurious leptons.  The sources of spurious electrons are
jet fluctuations and unresolved $e^+e^-$ pairs from photon conversions.
The probability of a jet faking an isolated muon is negligible.

The background from fake electrons was calculated from data using
dilepton events with one or more additional photons and/or jets.
The expected number of events was determined by multiplying the
number of events seen in data by the probability of a photon conversion or the
rate for a jet to fake an electron~\cite{SOSEBEE}.
The primary source of background in the $\mu\mu\mu$ channel is heavy flavor
($b\bar{b}$\ and $c\bar{c}$) events with the muons produced at large angle to
the jets. The total background for each final state is included in
Table~\ref{tab:AnalCuts}.

Based on zero candidate events, we present a 95\% confidence level upper limit
on the cross section for producing \wino$_1$\zino$_2$ pairs
times the branching ratio into any one of the trilepton final states. The
results from
the four channels were combined in the calculation of the limit, with the
assumption that
Br$(eee) =$ Br$(ee\mu) =$ Br$(e\mu\mu) =$ Br$(\mu\mu\mu)$. Uncertainties in
this calculation include the uncertainty in the luminosity (5.4\%)
and uncertainties in the overall analysis detection efficiencies (between
15\% and 25\% of the value) due to Monte Carlo statistics, systematic errors
in the determination of lepton identification efficiencies, systematic errors
in the trigger efficiencies, and systematic errors arising from energy scale
corrections. To construct this limit we used the Bayesian approach
of~\cite{PDG}, with the distribution of systematic errors represented by a
Gaussian and a flat prior probability distribution for the signal cross
section.

\begin{figure}
\centerline{\epsfxsize=6in \epsffile{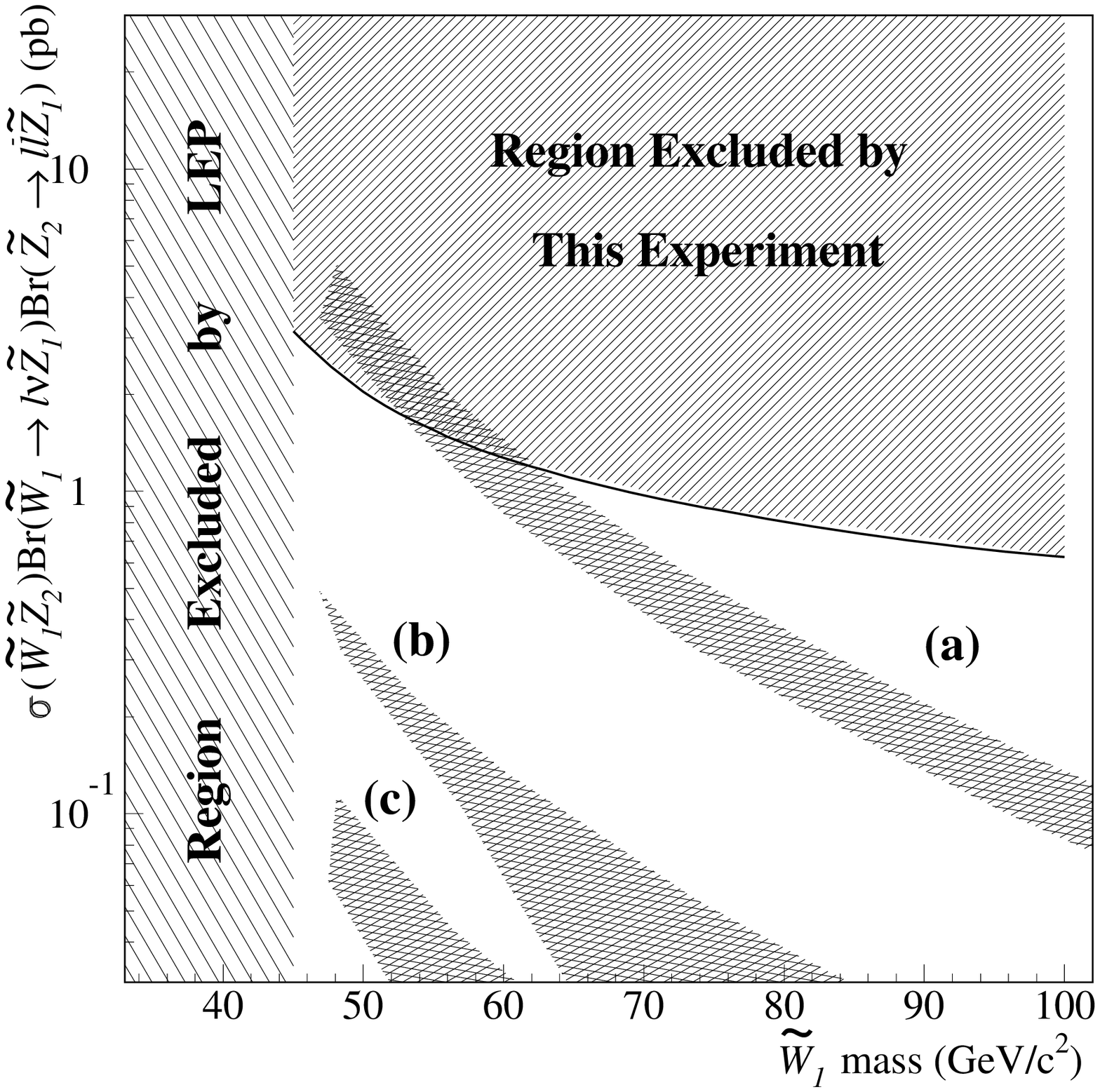}}
\vspace{6in}
\caption{The 95\% C.L. limit on cross section times branching
ratio into any one trilepton final state, as a function
of $M_{\widetilde{W}_1}$, along with the region of $M_{\widetilde{W}_1}$\
excluded by LEP. Also shown are bands of theoretical predictions, as described
in the text.}
\label{fig:limit}
\end{figure}

In Fig.~\ref{fig:limit} we show the resulting limit in the region above the LEP
limit~\cite{LEPLIMS}. For comparison, we show three bands of theoretical
curves. Band (a) shows the {\small ISAJET}
production cross section obtained with a wide range of input parameters,
multiplied by a
branching ratio of $1\over{9}$. The value of $1\over{9}$\ for a single
trilepton channel is obtained when the $\widetilde{W}_1$
and $\widetilde{Z}_2$\ decay purely leptonically and lepton universality is
applied. Branching ratios of this order are predicted in models with very light
sleptons, as for example the model of Ref.~\cite{LOPEZ}. Bands (b) and (c) show
the $\sigma\times$Br values from {\small ISAJET} obtained with the following
SUGRA input parameters:
$m_0 = [200,900]$\ \gevcc, $m_{1\over{2}} = [50,120]$\ \gevcc, $A_0 = 0$\ and
the sign of $\mu$\ negative. Band (b) is for $\tan{\beta} = 2$\ and band (c)
for $\tan{\beta} = 4$.

In conclusion, we have searched for the associated production of chargino and
neutralino pairs
by looking for the reaction
\ppbar$\ \to \widetilde{W}_1\widetilde{Z}_2 \to 3l + X$.
We see no evidence for \wino$_1$\zino$_2$\ production in 12.5 \ipb\ of data.
This leads to upper limits on
$\sigma(\widetilde{W}_1\widetilde{Z}_2)$Br$(\widetilde{W}_1\rightarrow l\nu
\widetilde{Z}_1)$Br$
(\widetilde{Z}_2\rightarrow l\bar{l}\widetilde{Z}_1)$\
ranging from 3.1 pb for $M_{\widetilde{W}_1} = 45$\ \gevcc\ to 0.6 pb for
$M_{\widetilde{W}_1} = 100$\ \gevcc.

%
We thank the Fermilab Accelerator, Computing, and Research Divisions, and
the support staffs at the collaborating institutions for their contributions
to the success of this work.   We also acknowledge the support of
the U.S. Department of Energy,
the U.S. National Science Foundation,
the Commissariat \`a L'Energie Atomique in France,
the Ministry for Atomic Energy and the Ministry of Science and Technology
   Policy in Russia,
CNPq in Brazil,
the Departments of Atomic Energy and Science and Education in India,
Colciencias in Colombia,
CONACyT in Mexico,
the Ministry of Education, Research Foundation and KOSEF in Korea,
CONICET and UBACYT in Argentina,
and the A.P. Sloan Foundation.
%

\small{
\begin{table}[htbp]
\begin{tabular}{cccccc}
\multicolumn{2}{c}{Channel}
                         & ~~$eee$~~  & ~~$ee\mu$~~
& ~~$e\mu\mu$~~ & ~~$\mu\mu\mu$~~   \\ \hline
          &              &              &
&               &               \\
\multicolumn{2}{c}{$\int{\cal{L}}$d$t$ (\ipb)}
                         & 12.5         & 12.5
& 12.2          &  10.8         \\  \hline
  ~~~~~~  &    ~~~~~
& \multicolumn{4}{c}{~~~~~~~~~{Events Remaining}~~~~~~~~~}     \\
\multicolumn{2}{c}{Cuts}

& \multicolumn{4}{c}{~~~~~~~~~{By Analysis Channel}~~~~~~~~~} \\ \hline
          &              &
&               &               &               \\
\multicolumn{2}{c}{$N_e + N_{\mu} \geq 3$}
                         &  13          &  42
& 297           &  2475         \\
\multicolumn{2}{c}{With Quality Cuts}
                         &   5          &      2
&   5           &     7         \\
\multicolumn{2}{c}{$N_{e}$ forward $< 2$}
                         &   4          &      0
&     N/A       &    N/A        \\
\multicolumn{2}{c}{ \met$ > 10$\ \gev}
                         &   0          &    N/A
&     N/A       &    N/A        \\
\multicolumn{2}{c}{$M_{\mu\mu} > 5$ \gevcc}
                         &   N/A        &    N/A
&   0           &     0         \\
          &              &              &
&               &               \\ \hline
\multicolumn{2}{c}{{\bf Candidates}}
                         &  0           &  0
&     0         &       0       \\ \hline
\multicolumn{2}{c}{Background}
                         & $0.8\pm 0.5$ & $0.8\pm 0.4$
& $0.6\pm 0.2$  & $0.1\pm 0.1$  \\
\end{tabular}
\vspace{0.2in}
\caption{Analysis cuts for each of the search channels, showing the number of
events left after a cut has been applied. No candidates are seen in any of the
four channels. The predicted background per channel is also shown.}
\label{tab:AnalCuts}
\end{table}
}
\normalsize

%
\end{document}